\documentclass{llncs}

\usepackage{amsfonts}

\begin{document}

\title{Small $\ell$-edge-covers in $k$-connected graphs}

\author{Zeev Nutov}
\institute{The Open University of Israel \ \email{nutov@openu.ac.il}}

\maketitle

\begin{abstract}
Let $G=(V,E)$ be a $k$-edge-connected graph with edge costs $\{c(e):e \in E\}$ and let $1 \leq \ell \leq k-1$. 
We show by a simple and short proof, that $G$ contains an $\ell$-edge cover $I$ such that:
$c(I) \leq \frac{\ell}{k}c(E)$ if $G$ is bipartite, or 
if $\ell |V|$ is even, or 
if $|E| \geq \frac{k|V|}{2} +\frac{k}{2\ell}$; otherwise, 
$c(I) 
\leq \left(\frac{\ell}{k}+\frac{1}{k|V|}\right)c(E)$.
The particular case $\ell=k-1$ and unit costs already includes a result of Cheriyan and Thurimella \cite{CT}, 
that $G$ contains a $(k-1)$-edge-cover of size $|E|-\lfloor |V|/2 \rfloor$.
Using our result, we slightly improve the approximation ratios for 
the {\sf $k$-Connected Subgraph} problem (the node-connectivity version) 
with uniform and $\beta$-metric costs.
We then consider the dual problem of finding a spanning subgraph of maxi\-mum connectivity $k^*$ 
with a prescribed number of edges. We give an algorithm that computes a $(k^*-1)$-connected subgraph,
which is tight, since the problem is NP-hard.  
\end{abstract}


\section{Introduction}

Let $G=(V,E)$ be an undirected graph, possibly with parallel edges. For $S \subseteq V$ let 
$\delta(S)$ denote the set of edges in $E$ with exactly one endnode in $S$.
Let $n=|V|$.
An edge set $I \subseteq E$ is an {\em $\ell$-edge-cover} (of  $V$) if the graph $(V,I)$ has 
minimum degree $\geq \ell$.
For $x \in \mathbb{R}^E$ and $F \subseteq E$ let $x(F)=\sum_{e \in F} x(e)$.
Let $P^f_{cov}(G,\ell)$ denote the {\em fractional $\ell$-edge-cover polytope} 
determined by the linear constraints
\begin{eqnarray}
\label{e:1}  x(\delta(v))                      & \geq & \ell                               
\hphantom{aaaaaaaaaaa}        v \in V  \\
\nonumber 1 \geq x_e                & \geq & 0    \hphantom{aaaaaaaaaaa}    e \in E 
\end{eqnarray}
Let $P_{cov}(G,\ell)$ denote the {\em integral $\ell$-edge-cover polytope}, 
which is the convex hull of the characteristic vectors of of the $\ell$-edge-covers in $G$. 
It is known that if $G$ is bipartite then $P^f_{cov}(G,\ell)=P_{cov}(G,\ell)$
(see \cite{Sch}, (31.7) on page 340). This implies the following.

\begin{proposition} \label{p:cov}
Let $G=(V,E)$ be a bipartite graph and let $1 \leq \ell \leq k-1$.
Let $x \in P^f_{cov}(G,k)$. Then $\frac{\ell}{k} x \in P_{cov}(G,\ell)$.
\end{proposition}

\begin{corollary} \label{c:cov}
Let $G=(V,E)$ be a bipartite graph with 
edge costs $\{c(e):e \in E\}$ and minimum degree $\geq k \geq 2$.
Then for any $1 \leq \ell \leq k-1$ \ $G$ contains an $\ell$-edge cover $I \subseteq E$ of cost 
$c(I) \leq \frac{\ell}{k}c(E)$.
\end{corollary}

Cheriyan an Thurimella \cite{CT} showed that if $G$ is bipartite and has minimum degree $\geq k$, 
then $G$ contains a $(k-1)$-edge-cover $I$ such that $|I| \leq |E|-n/2$.
Note that this bound follows from Corollary~\ref{c:cov} by 
assuming unit costs,
substituting $\ell=k-1$, and 
observing that $|E| \geq \frac{kn}{2}$.
Unfortunately, Corollary~\ref{c:cov} does not extend to the 
general (non-bipartite) case, 
e.g., if $G$ is a cycle of length $3$, $k=2$, and $\ell=1$. 
On the positive side, it is proved in
\cite{DN} that if $G$ has minimum degree $\geq k$ then $G$ contains 
a $(k-1)$-edge-cover $I$ of cost $c(I) \leq \frac{2k-2}{2k-1} c(E)$.
Let $\zeta(S)$ denote the set of edges in $E$ with at least one endnode in $S$.
It is known that in the general case, 
$P_{cov}(G,\ell)$ is determined by adding to the constraints 
of $P^f_{cov}(G,\ell)$ the following inequalities (see \cite{Sch}, page 581, Theorem 34.13)
\begin{equation} \label{e:2} 
x(\zeta(S) \setminus F) \geq  \frac{\ell|S|}{2}-\frac{|F|-1}{2} \ \ \ \     
S \subseteq V, F \subseteq \delta(S), \ell|S|-|F| \geq 1 \mbox{ odd } \ . 
\end{equation}

A graph $G$ is {\em $k$-edge-connected} if $|\delta(S)| \geq k$ for all $\emptyset \neq S \subset V$.
Cheriyan and Thurimella \cite{CT} showed that if $G$ is $k$-edge-connected, 
then $G$ contains a $(k-1)$-edge-cover $I$ such that $|I| \leq |E|-\lfloor n/2 \rfloor$.
We present an analogue of Proposition~\ref{p:cov} and Corollary~\ref{c:cov} for general graphs,
with simple and short proof, that also implies this bound of \cite{CT}.
Let $P^f_{con}(G,k)$ denote the {\em fractional $k$-edge-connectivity polytope}, 
determined by 
\begin{eqnarray}
\nonumber x(\delta(S))                      & \geq & k                               
\hphantom{aaaaaaaa}        \emptyset \neq S \subset V  \\
\nonumber 1 \geq x_e                & \geq & 0    \hphantom{aaaaaaaa}    e \in E 
\end{eqnarray}
Note that $P^f_{cov}(G,k) \subseteq P^f_{con}(G,k)$, and that if $x \in P^f_{cov}(G,k)$ then $x(E) \geq \frac{kn}{2}$. 
The main result of this paper is the following analogue of Proposition~\ref{p:cov}.

\begin{theorem} \label{t:main}
Let $G=(V,E)$ be a graph, let $1 \leq \ell \leq k-1$, and let~$x \in P^f_{con}(G,k)$.
Then $\frac{\ell}{k} x \in P_{cov}(G,\ell)$ if $\ell |V|$ is even or if $x(E) \geq \frac{k|V|}{2} +\frac{k}{2\ell}$;
otherwise, $\frac{\ell|V|+1}{2x(E)} \cdot x \in P_{cov}(G,\ell)$, and hence also 
$\left(\frac{\ell}{k}+\frac{1}{k|V|}\right) \cdot x \in P_{cov}(G,\ell)$.
\end{theorem}

Theorem~\ref{t:main} immediately implies the following. 

\begin{corollary} \label{c:CT}
Let $G=(V,E)$ be a $k$-edge-connected graph with edge costs $\{c(e):e \in E\}$ and
let $1 \leq \ell \leq k-1$. 
Then $G$ contains an $\ell$-edge cover $I \subseteq E$ such that: 
$c(I) \leq \frac{\ell}{k} c(E)$ if $\ell |V|$ is even or if $|E| \geq \frac{kn}{2} +\frac{k}{2\ell}$;
otherwise, $c(I) \leq \frac{\ell|V|+1}{2|E|}c(E) \leq \left(\frac{\ell}{k}+\frac{1}{k|V|}\right)c(E)$.
\end{corollary}

Note that the bound $|I| \leq |E|-\lfloor n/2 \rfloor$ of Cheriyan and Thurimella \cite{CT} 
follows from Corollary~\ref{c:CT} by 
assuming unit costs,
substituting $\ell=k-1$, and 
observing that $|E| \geq \frac{kn}{2}$.
Indeed, by Corollary~\ref{c:CT}, $|E|-|I| \geq |E|/k \geq n/2$
if $(k-1)n$ is even or if $|E| \geq \frac{kn}{2}+1$.
Otherwise, $k$ is even, $n$ is odd, $|E|=\frac{kn}{2}$, 
and then, by Corollary~\ref{c:CT}, $|E|-|I| \geq \frac{n-1}{kn} |E| = \frac{n-1}{2}= \lfloor n/2 \rfloor$.

We now discuss some applications of Corollaries \ref{c:cov} and \ref{c:CT} for both directed and undirected graphs, 
for the following classic NP-hard problem.
A (simple) directed or undirected graph is {\em $k$-connected} if it contains $k$ internally disjoint 
paths from every node to the other. 

\begin{center} 
\fbox{
\begin{minipage}{0.960\textwidth}
\noindent
{\sf $k$-Connected Subgraph} \\
{\em Instance:} \ 
A graph $G'=(V,E')$ with edge costs and an integer $k$. \\
{\em Objective:} 
Find a minimum cost $k$-connected spanning subgraph $G$ of $G'$. 
\end{minipage}
}
\end{center}

The case of unit costs is the {\sf Minimum Size $k$-Connected Subgraph} problem.
Cheriyan and Thurimella \cite{CT} suggested and analyzed the following algorithm 
for the {\sf Minimum Size $k$-Connected Subgraph} problem, for both directed and undirected graphs;
in the case of a directed graph $G=(V,E)$, we say that 
$I \subseteq E$ is an $\ell$-edge-cover if $(V,I)$ has minimum outdegree 
and minimum indegree $\geq \ell$. 

\begin{center} 
\fbox{
\begin{minipage}{0.960\textwidth}
\noindent
{\bf Algorithm 1}
\begin{enumerate}
\item
Find a minimum size $(k-1)$-edge cover $I \subseteq E$.
\item
Find an inclusion minimal edge set $F \subseteq E \setminus I$ such that $(V,I \cup F)$ is $k$-connected.
\item
Return $I \cup F$.
\end{enumerate}
\end{minipage}
}
\end{center}

They showed that this algorithm has approximation ratios
\begin{itemize}
\item[$\bullet$]
$1+\frac{n}{\sf opt}  \leq 1+\frac{1}{k}$ \ for directed graphs;  
\item[$\bullet$]
$1+\frac{n}{2{\sf opt}} \leq 1+\frac{1}{k}$ for undirected graphs.
\end{itemize}
Here ${\sf opt}$ denotes the optimum solution value of a problem instance at hand. 
Step~1 in the algorithm can be implemented in polynomial time, c.f. \cite {Sch}.
Recently, the performance of this algorithm was also analyzed in \cite{DN} for so called 
{\em $\beta$-metric costs}, when the input graph is complete and
for some $1/2 \leq \beta < 1$ the costs satisfy the {\em $\beta$-triangle inequality} 
$c(uv) \leq \beta [c(ua)+c(av)]$ 
for all $u,a,v \in V$. When $\beta=1/2$, the costs are uniform, and we have the min-size version of 
the problem. If we allow the case $\beta= 1$, then the costs satisfy the ordinary
triangle inequality and we have the metric version of the problem. 
In \cite{DN} it is shown that for undirected graphs with $\beta$-metric costs the above algorithm has ratio
$1-\frac{1}{2k-1}+\frac{2\beta}{k(1-\beta)}$.
We prove the following. 

\begin{theorem} \label{t:ratio}
\begin{itemize}
\item[{\em (i)}]
For the {\sf Minimum Size $k$-Connected Subgraph} problem, Algorithm~1 has approximation ratios
\begin{itemize}
\item 
$1-\frac{1}{k}+2n/{\sf opt}  \leq 1+\frac{n}{\sf opt}$ for directed graphs;  
\item
$1-\frac{1}{k}+n/{\sf opt} \leq 1+\frac{n}{2 \sf opt}$ \ for undirected graphs.
\end{itemize}
\item[{\em (ii)}]
In the case of undirected graphs and $\beta$-metric costs, Algorithm~1 has
approximation ratio 
$1-\frac{1}{k}+\frac{1}{kn}+ \frac{2\beta}{k(1-\beta)}$.
\item[{\em (iii)}]
There exists a polynomial time algorithm that given an instance of the {\sf Minimum Size $k$-Connected Subgraph}
problem returns a $(k-1)$-connected spanning subgraph $G$ of $G'$ with at most ${\sf opt}$ edges.
\end{itemize}
\end{theorem}

Note that in part~(i) of Theorem~\ref{t:ratio} we do {\em not} improve the worse performance guarantee 
$1+\frac{1}{k}$ of \cite{CT}.
However, the ratio $1+\frac{1}{k}$ applies only if ${\sf opt}=kn$ in the case of directed graphs
and ${\sf opt}=kn/2$ in the case of undirected graphs. Otherwise, if ${\sf opt}$ is larger 
than these minimum possible values, 
then both our analysis and that of \cite{CT} give better ratios. 
But the ratios provided by our analysis are smaller, since 
$2n/{\sf opt}-\frac{1}{k} \leq n/{\sf opt}$ in the case of directed graphs, and
$n/{\sf opt}-\frac{1}{k} \leq n/2{\sf opt}$ in the case of undirected graphs.
For example, in the case of directed graphs, 
if ${\sf opt}=\frac{3}{2}kn$ then our ratio is $1+\frac{1}{3k}$, while that of \cite{CT} is $1+\frac{2}{3k}$.


Part~(iii) of Theorem~\ref{t:ratio} can be used to obtain a tight approximation algorithm
to the {\sf Maximum Connectivity $m$-Edge Subgraph} problem:
given a graph $G'$ and an integer $m$, find a spanning subgraph $G$ of $G'$ with at most $m$ edges
and maximum connectivity $k^*$.
We can apply the algorithm in part~(iii) to find the maximum integer $k$ for which the 
algorithm returns a subgraph with at most $m$ edges. 
Then $k \geq k^*-1$, hence we obtain a 
polynomial time algorithm that computes a 
$(k^*-1)$-connected spanning subgraph with at most $m$ edges. 
Note that this is tight, since the problem is NP-hard. 

\section{Proof of Theorem~\ref{t:main}}

Let $x \in P^f_{con}(G,k)$ and let $S \subseteq V$. 
It is clear that inequalities (\ref{e:1}) are ``scalable'' by $\frac{\ell}{k}$, namely,
$\frac{\ell}{k} x(\delta(S)) \geq \ell$. We show that inequalities (\ref{e:2}) are also ``scalable''
by a factor of $\mu$ defined as follows: 
$\mu=\frac{\ell}{k}$ if $\ell n$ is even or if $x(E) \geq \frac{kn}{2} +\frac{k}{2\ell}$,
and $\mu=\frac{\ell n+1}{2x(E)}$ otherwise.
Let $F \subseteq \delta(S)$ such that $\ell|S|-|F| \geq 1$ is odd. 
We prove that then the following holds:
\begin{equation} \label{e:kk} 
\mu (x(\zeta(S))-x(F)) \geq  \frac{\ell|S|}{2}-\frac{|F|-1}{2} \ .
\end{equation}

If $S=V$ then $\zeta(S)=E$ and $F=\emptyset$. Then (\ref{e:kk}) reduces to a void condition if $\ell|V|$ is even, 
and to the condition $\mu x(E) \geq \frac{\ell n+1}{2}$ otherwise, which holds by the definition of $\mu$.

Henceforth assume that $S$ is a proper subset of $V$.
We prove that then 

\begin{equation} \label{e:k-1} 
\frac{\ell}{k} (x(\zeta(S))-x(F)) \geq  \frac{\ell|S|}{2}-\frac{|F|-1}{2} \ .
\end{equation}

Multiplying both sides of (\ref{e:k-1}) by $\frac{k}{\ell}$ gives 
\begin{equation} \label{e:k-1'} 
x(\zeta(S))-x(F) \geq \frac{k|S|}{2}-\frac{k}{\ell} \cdot \frac{|F|-1}{2} \ .
\end{equation}

Note that 
$x(\zeta(S)) \geq \frac{k|S|}{2}+\frac{x(\delta(S))}{2}$ and that $x(F) \leq |F|$.
Substituting and rearranging terms, we obtain that it is sufficient to prove that 
if $x(\delta(S)) \geq k \geq \ell+1 \geq 0$, then 
\begin{equation} \label{e:final}
x(\delta(S))-x(F) +\frac{k-\ell}{\ell}|F| \geq \frac{k}{\ell} \ \ \ \ \ \emptyset \neq S \subset V \ .
\end{equation}
If $|F| \geq \frac{k}{k-\ell}$ then (\ref{e:final}) holds, since $x(\delta(S)) \geq x(F)$.
Assume that $|F| < \frac{k}{k-\ell}$. Then
$$
x(\delta(S))-x(F) +\frac{k-\ell}{\ell}|F| \geq k-|F|+\frac{k-\ell}{\ell}|F|= k+\frac{k-2\ell}{\ell}|F| \geq \frac{k}{\ell} \ .
$$
We explain the last inequality. 
If $k \geq 2\ell$ then $k+\frac{k-2\ell}{\ell}|F| \geq k \geq \frac{k}{\ell}$.
If $k < 2\ell$ then since $|F| < \frac{k}{k-\ell}$
$$k+\frac{k-2\ell}{\ell}|F| > k+\frac{k-2\ell}{\ell}\cdot \frac{k}{k-\ell}= 
k+\frac{k}{\ell}-\frac{k}{k-\ell} \geq \frac{k}{\ell} \ .$$

In both cases (\ref{e:final}) holds, and hence the proof of Theorem~\ref{t:main} is complete.

\section{Proof of Theorem~\ref{t:ratio}}

Let $I$ and $F$ denote the set of edges computed by Algorithm~1 
at steps 1 and 2, respectively.
We prove part (i), starting with the case of directed graphs.
For a directed graph $G$, the corresponding bipartite graph $G'=(V \cup V',E')$ is obtained by 
adding a copy $V'$ of $V$ and replacing every directed edge $uv \in E$ by the undirected edge $uv'$,
where $v' \in V'$ is the copy of $v$. 
It is not hard to verify that $I$ is an $\ell$-edge-cover in $G$ if, and only if, 
the set $I'$ of edges that corresponds to $I$ is an $\ell$-edge-cover in $G'$.
Thus $|I| \leq \frac{k-1}{k} {\sf opt}$, by Corollary~\ref{c:cov}.
On the other hand, by the directed Critical Cycle Theorem of Mader \cite{Mad-dir} (see \cite{CT} for details), 
the set of edges of $G'$ that corresponds to $F'$ forms a forest in $G'$, hence $|F| \leq 2n-1$. Consequently,
$\frac{|I|+|F|}{\sf opt} \leq 1-\frac{1}{k} +\frac{2n-1}{\sf opt}$.

Let us consider undirected graphs.
If $(k-1)n$ is even or if ${\sf opt} \geq \frac{kn}{2} +\frac{k}{2(k-1)} \geq \frac{kn}{2} +1$,
then $|I| \leq \frac{k-1}{k} {\sf opt}$, by Corollary~\ref{c:CT}.
By the undirected Critical Cycle Theorem of Mader \cite{Mad-und} (see \cite{CT} for details),
$F$ is a forest, hence $|F| \leq n-1$. Consequently,
$\frac{|I|+|F|}{\sf opt} \leq 1-\frac{1}{k} +\frac{n-1}{\sf opt}$.
If $(k-1)n$ is odd and ${\sf opt} < \frac{kn}{2} +1$, then an optimal solution is $k$-regular
and hence $|I| \leq \frac{(k-1)n+1}{2} \leq \left(1-\frac{1}{k}\right)({\sf opt}+1)$.
Combining we get 
$\frac{|I|+|F|}{\sf opt} \leq 1-\frac{1}{k} + \frac{1-1/k}{\sf opt} + \frac{n-1}{\sf opt} 
< 1-\frac{1}{k} + \frac{n}{\sf opt}$.

Now let us consider part~(ii), the case of $\beta$-metric costs. 
In \cite{DN} it is proved that $c(F) \leq \frac{2\beta}{k(1-\beta)} {\sf opt}$.
If $(k-1)n$ is even, or if there exists an optimal solution with at least 
$\frac{kn}{2}+\frac{k}{2(k-1)} \leq \frac{kn}{2}+1$ edges, 
then Corollary~\ref{c:CT} gives the bound $c(I) \leq \left(1-\frac{1}{k}\right){\sf opt}$. 
Else, Corollary~\ref{c:CT} gives the bound 
$c(I) \leq \left(1-\frac{1}{k}+\frac{1}{kn}\right){\sf opt}$, and the result follows.

We prove part~(iii).
We apply Algorithm~1 with $k$ replaced by $k-1$, namely,
$I \subseteq E$ is a minimum size $(k-2)$-edge cover and 
$F \subseteq E \setminus I$ is an inclusion minimal edge set 
such that $(V,I \cup F)$ is $(k-1)$-connected.
Now we use the bounds in Corollary~\ref{c:CT}.
In the case of directed graphs we have $|I| \leq \frac{k-2}{k} {\sf opt}$,
$|F| \leq 2n-1 \leq \frac{2}{k} {\sf opt}$, and the result follows.
In the case of undirected graphs we have $|I| \leq \left(\frac{k-2}{k}+\frac{1}{kn}\right) {\sf opt}$ and
$|F| \leq n-1 \leq \left(\frac{2}{k} - \frac{2}{kn} \right){\sf opt}$, and the result follows.

The proof of Theorem~\ref{t:ratio} is complete.


\begin{thebibliography}{1}

\bibitem{CT}
J.~Cheriyan and R.~Thurimella.
\newblock Approximating minimum-size $k$-connected spanning subgraphs via
  matching.
\newblock {\em SIAM J. Computing}, 30:528--560, 2000.

\bibitem{DN}
J.~David and Z.~Nutov.
\newblock Approximating survivable networks with $\beta$-metric costs.
\newblock {\em J. Discrete Algorithms}, 9(2):170--175, 2011.

\bibitem{Mad-und}
W.~Mader.
\newblock Ecken vom grad $n$ in minimalen n-fach zusammenh\"{a}ngenden graphen.
\newblock {\em Archive der Mathematik}, 23:219--224, 1972.

\bibitem{Mad-dir}
W.~Mader.
\newblock Minimal $n$-fach in minimalen n-fach zusammenh\"{a}ngenden digraphen.
\newblock {\em J. Comb. Theory B}, 38:102--117, 1985.

\bibitem{Sch}
A.~Schrijver.
\newblock {\em Combinatorial Optimization Polyhedra and Efficiency}.
\newblock Springer, 2004.

\end{thebibliography}


\end{document}